\documentclass{PoS}

\title{Quark Recombination in Heavy Ion Collisions}

\ShortTitle{Quark Recombination}

\author{\speaker{Rainer J.\ Fries}\\ %\thanks{A footnote may follow.}\\
        Cyclotron Institute and Department of Physics and Astronomy, Texas
        A\&M University, College Station TX, 77845, USA\\
        RIKEN/BNL Research Center, Brookhaven National Laboratory, Upton NY,
        11973, USA\\
        E-mail: \email{rjfries@comp.tamu.edu}}

\abstract{Data on high energy nuclear collisions collected at the Relativistic
Heavy Ion Collider over the past decade have provided convincing evidence
that hadronization is quite different in hot nuclear environments compared to
$p+p$ collisions. In particular, the data suggest that we see traces
of quark degrees of freedom in elliptic flow, with the implication that
collective flow is generated on the parton level and is transfered to hadrons
through a simple recombination step. In this contribution we review the
experimental evidence for quark recombination and discuss some
recombination models which are used to describe these effects. }

\FullConference{Workshop on Critical Examination of RHIC Paradigms - CERP2010\\
		April 14-17, 2010\\
		Austin Texas USA}

\begin{document}

\section{Introduction}

The experimental program at the Relativistic Heavy Ion Collider (RHIC) was
born from the idea that a deconfined state of quarks and gluons, the quark
gluon plasma (QGP), can be created and studied in collisions of heavy nuclei at the
highest possible collision energies. Soon after turning on RHIC it became
clear that we indeed see novel and unusual phenomena \cite{whitepaper:05}. 
Some of those like strong jet
quenching had been predicted qualitatively, others, like the quark number
scaling of elliptic flow were surprising. Over the years a consensus seems to
have emerged that quarks and gluons are indeed deconfined for a short amount
of time in the
fireball created at RHIC, and that this quark gluon plasma behaves like a very
good liquid with small viscosity over entropy ratio $\eta/s$. The key ingredient
for this conclusion was the comparison of data with ideal (and later viscous)
hydrodynamic calculations based on equations of state with partonic phases
and small $\eta/s$ \cite{whitepaper:05,Gyulassy:2004zy}. However, there are 
plenty of reasons to check whether alternative explanations can be 
ruled out with certainty.

In this contribution we review the role of quark recombination models at RHIC
\cite{Becattini:2009fv,Fries:2008hs,Fries:2006iv,Fries:2004ej}.
The main motivation for their emergence was the anomalous baryon enhancement
and the observed quark number scaling 
of elliptic flow \cite{Adler:2003kg,Adams:2003am,Adams:2004bi} . Such 
recombination models provide strong
evidence that collective flow is partonic in origin. In particular, quarks
and gluons seem to be the relevant degrees of freedom when elliptic flow is built up.
This touches a topic that is of interest beyond heavy ion physics,
namely the still not understood phenomenon of hadronization.

Hadronization, i.e.\ the color neutralization process that requires quarks and
gluons to form hadrons, is a non-perturbative phenomenon that has largely
defied a first-principle computation. There are 3 basic approaches to
hadronization that are relevant in our context:
\begin{itemize}
\item Factorization: In some cases with large momentum transfer hadronization
  can be separated from the underlying (scattering) process in a rigorous way 
  \cite{Collins:1989gx}. This works for single quarks and
  gluons fragmenting into jets in the vacuum at sufficiently large
  momentum and for exclusive processes at large momentum transfer. 
  The fragmentation process for jets is universal and can be parameterized through 
  fragmentation functions \cite{Albino:2008gy}.
\item Statistical and cluster emission concepts: They can do very well explaining
  certain bulk features of hadron production like hadron ratios, 
  see e.g.\ \cite{Becattini:2009fv}.
\item Microscopic models: They try to capture certain aspects of the
  underlying microscopic dynamics, though they are not comprehensive or 
  derived from first principles. Examples are string fragmentation or quark recombination. 
\end{itemize}

\begin{figure}[t]
\centerline{
%\includegraphics[width=5.0cm]{Jia_Mehen_reco.eps}
%\hspace{-0.2cm}
\includegraphics[width=7.0cm,angle=-90]{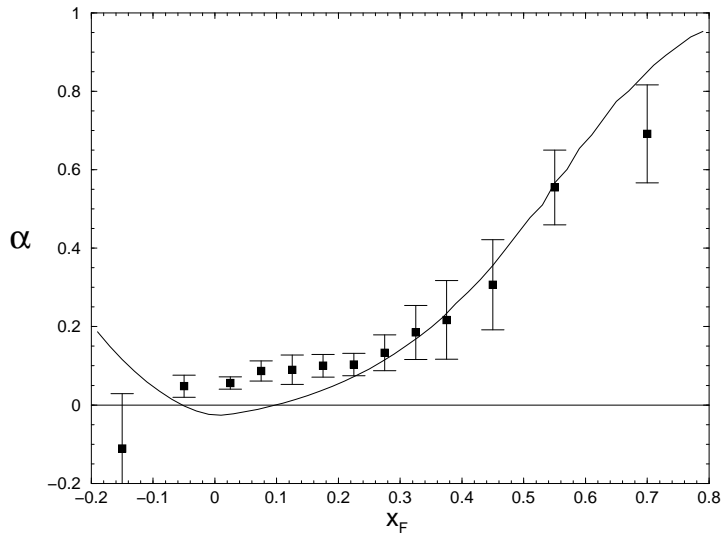}}
\caption{Asymmetry of negative and positive $D$ mesons as a function of 
  $x_F$ after Jia, Mehen and Braaten \cite{Braaten:2002yt} with data from 
  E791 \cite{Aitala:1996hf}.}
\label{fig:1}
\end{figure}

The idea that quarks coalesce into bound states similar to the coalescence of
nucleons into light nuclei or plasma constituents into atoms has been around since the
beginning of quark models and quantum chromodynamics (QCD) 
\cite{Das:1977cp,Gupt:1983rq,Lopez:1984,Biro:1994mp}. The
nature of QCD as a non-linear relativistic quantum field theory with a very
complex non-perturbative sector limited the success of recombination models
to particular situations. Generally those are characterized by the feature
that a well-defined multi-quark state for hadronization can be identified.
This is particularly true for the leading particle effect, in which a quark
(heavy flavors such as charm and strange quarks are experimentally
accessible using identified particle tags) is produced in a collision in 
forward direction and coalesces with a quark from the beam remnant. 
One example is the $D^-/D^+$ asymmetry observed in the fixed target experiment
E791 which used a $\pi^-$ beam on a nuclear target \cite{Aitala:1996hf}. The
asymmetry which grows to almost 100\% at extreme forward direction 
(Feynman-$x_F \to 1$) can be explained by
$c\bar c$ pair production with a preferential recombination of the $\bar c$
with the $d$ valence quark from the pion fragments, while the corresponding
$c+ \bar d$ combination does not involve a valence quark of the pion and is thus 
suppressed, see Fig.\ \ref{fig:1}.  The leading particle effect is probably the most convincing
argument for the existence of a quark recombination mechanism outside of
heavy ion physics \cite{Braaten:2002yt}.

\section{Experimental Evidence}

\begin{figure}[t]
\centerline{
\includegraphics[width=14.0cm]{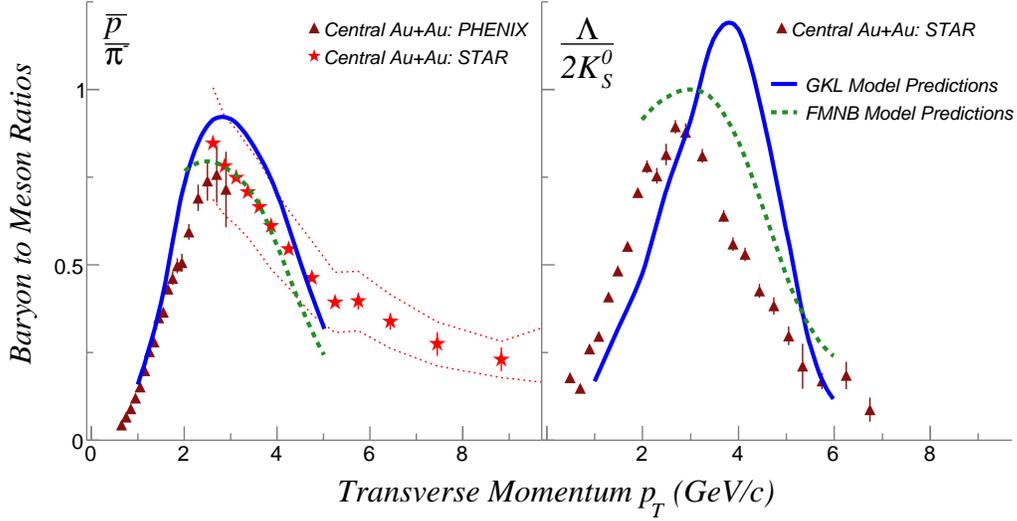}
%\hspace{-0.2cm}
%\includegraphics[width=8.0cm]{Rcp_summary.eps}
}
\caption{Data from STAR and PHENIX on $p/\pi$ and 
$K/\Lambda$ ratios in central Au+Au collisions at RHIC together with model 
calculations in the GKL and FMNB formalisms \cite{Fries:2008hs}.} 
\label{fig:2}
\end{figure}

\begin{figure}[t]
\centerline{
\includegraphics[width=8.0cm]{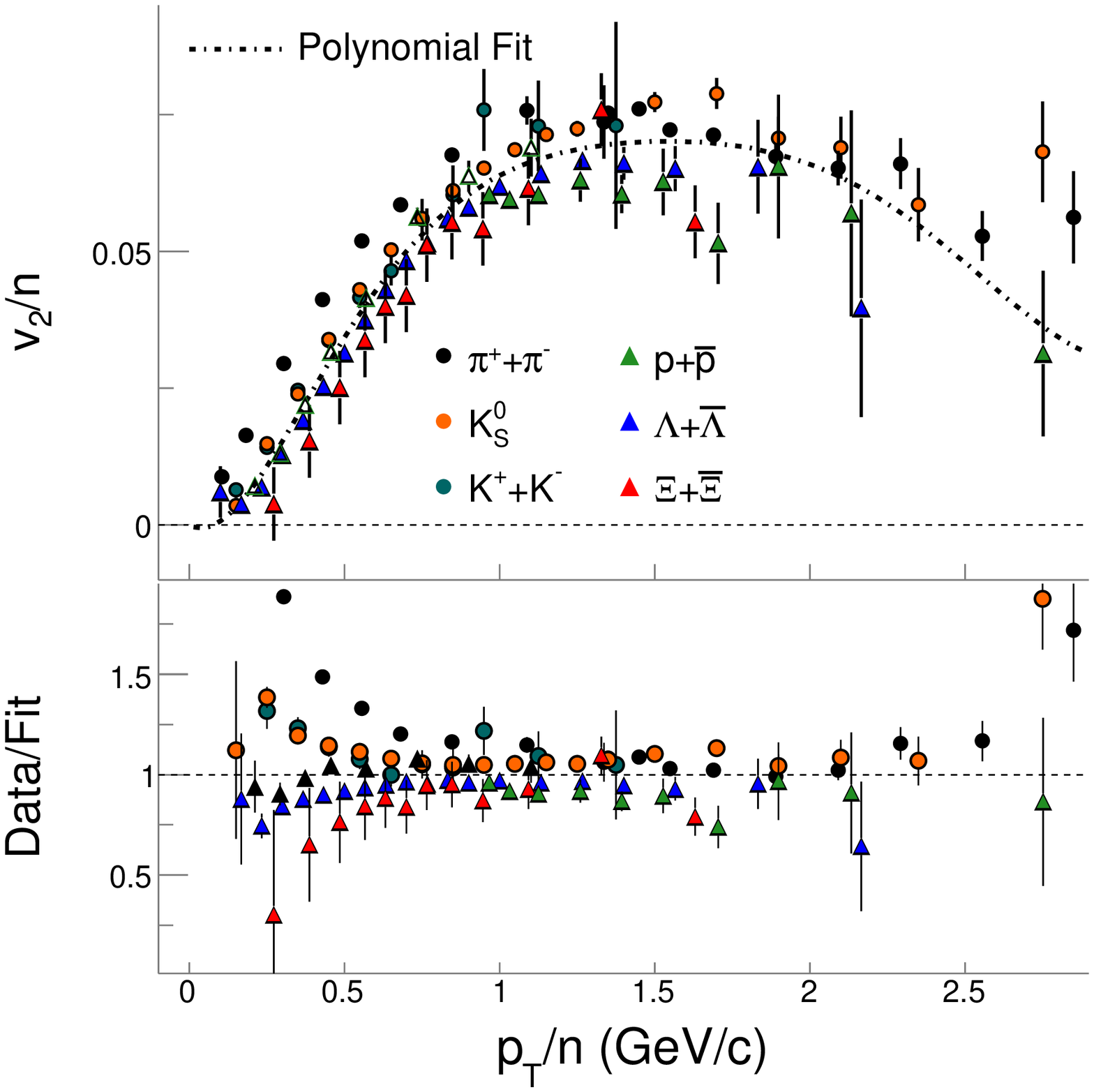}
%\hspace{-0.2cm}
\includegraphics[width=8.0cm]{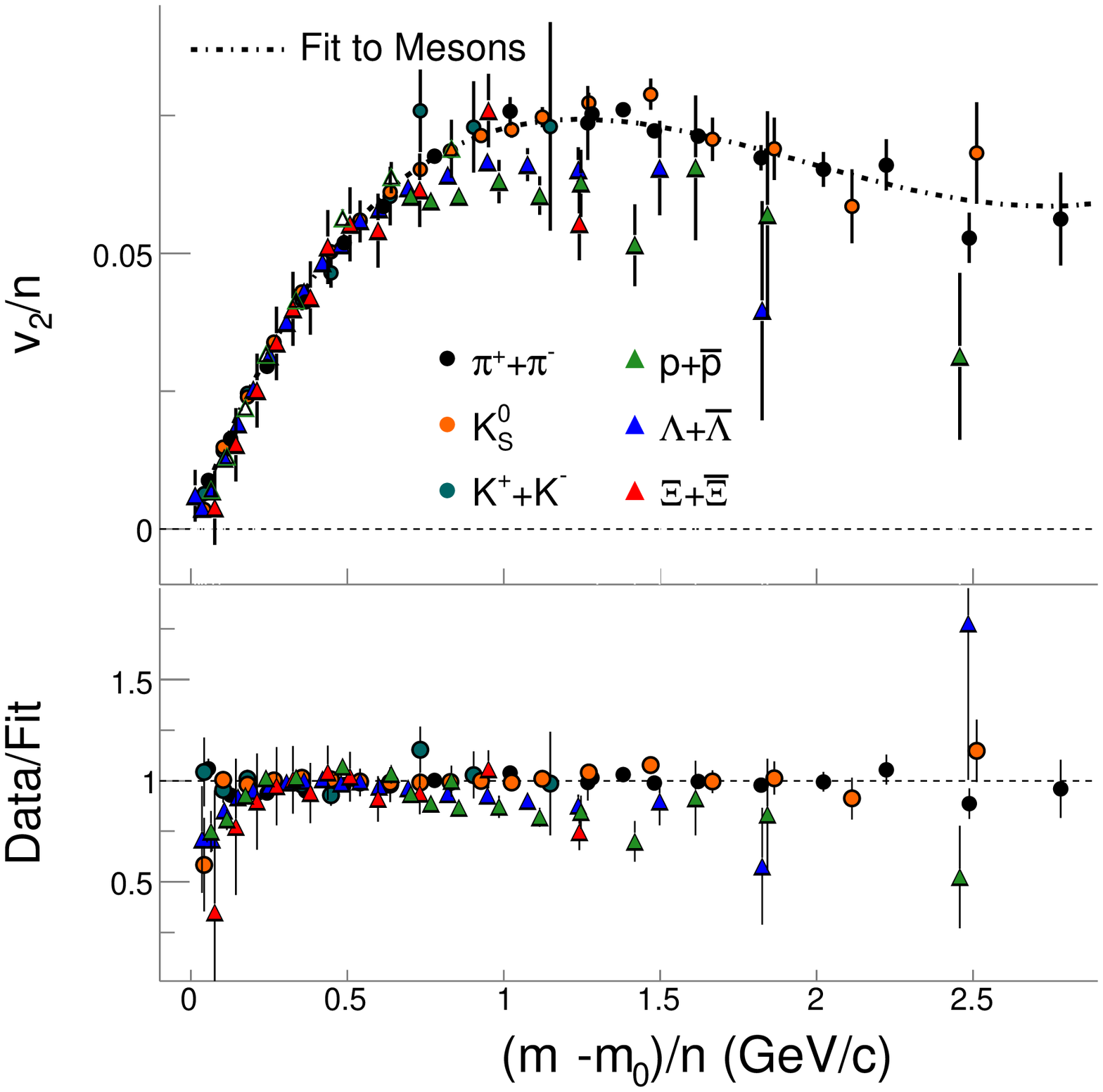}}
\caption{Left panel:  Quark number-scaled elliptic flow $v_2$ as a function of
  quark number-scaled transverse momentum (top) and deviations from a center 
  fit (bottom) to demonstrate the accuracy of the scaling behavior.
  Right panel: the same plotted vs quark number-scaled kinetic energy $m_T-m$.
  The scaling at low momenta or kinetic energies is improved in this case 
  \cite{Fries:2008hs}.}
\label{fig:3}
\end{figure}

A factor 5 suppression of high momentum hadrons was found soon after RHIC was 
turned on. This jet quenching phenomenon was expected from the predictions of
parton energy loss. The suppression was roughly consistent for pions and kaons
with a few GeV/$c$ transverse momentum $P_T$. However, the discovery that 
protons and $\Lambda$ baryons show little or no suppression was a surprise.
It also threatened the partonic interpretation of energy loss since jet
hadronization, even if altered by the presence of a medium was thought to
basically transfer quenching equally to all hadrons. These findings, that
are now the cornerstones
of experimental evidence for quark recombination, were first known as the
baryon anomaly at RHIC, since they defied our expectations for the intermediate
$P_T$ range. In that $P_T$ range (roughly between 1 and 5 GeV/$c$)
we find \cite{whitepaper:05,Gyulassy:2004zy}. 
\begin{itemize}
\item anomalously large baryon-to-meson ratios which were up to a factor 4
  larger than expected from $e^++e^-$ or $p+p$ collisions, see Fig.\ \ref{fig:2}. The
  proton-to-pion ratio can be one.
\item systematically larger suppression (shown by smaller nuclear modification
  factors $R_{AA}$ and $R_{CP}$) for mesons than for baryons.
\item systematically larger elliptic flow for baryons than for mesons with peak
  values roughly 50\% larger. This was later recognized to follow the simple
  scaling law (see Fig.\ \ref{fig:3})
\begin{equation}
  \frac{1}{3}  v_2^B (3 P_T) = \frac{1}{2} v_2^M (2 P_T)  
  \label{eq:scaling}
\end{equation}
where the factors 3 and 2 refer to the number of valence quarks in baryons
 $B$ and mesons $M$ respectively \cite{Voloshin:2002wa}.
\end{itemize}

Prior to RHIC this intermediate $P_T$ regime was expected to be dominated by
the physics of jets and hard processes, but the experimental data seemed to be
telling otherwise. Attempts to treat this as a transition region between bulk
hydrodynamics below 1 to 2 GeV/$c$ and pure jet quenching and fragmentation
at larger $P_T$ failed to capture crucial details, e.g.\ the fact that the
$\phi$ meson does not behave similar to the almost equally heavy proton, but
rather than the much lighter pion \cite{Hirano:2003pw}. This is evidence that at 
intermediate $P_T$
the number of valence quarks, and hence hadronization, is more important than 
collectivity in the hadronic phase which in a hydrodynamic picture depends
solely on the mass of the hadron. We conclude that the intermediate $P_T$
region in heavy ion collisions shows features which are neither hydrodynamic 
nor consequences of jet fragmentation.

The basic experimental findings of the early RHIC years have stood the test
of time. In more recent  years it has been found that a quark number-scaling of
$v_2$ using kinetic energy instead of transverse momentum improves the scaling
at low momentum (or kinetic energy), see Fig.\ \ref{fig:3}. But as we will
argue below this has no direct implications for quark recombination.

\section{Modeling Quark Recombination}

As in the case of the leading particle effect, recombination models for heavy
ion collisions are based
on the notion of a well-defined distribution of quarks just before
hadronization. This is thought to be a thermalized
plasma characterized by a temperature $T = T_c +\epsilon$ with some modest 
deviations from equilibrium allowed. Usually it is never clearly specified
what the precise assumptions are, but the following properties seem to be
important in most models:
\begin{itemize}
\item Gluons are frozen as degrees of freedom and quarks have already acquired 
  effective constituent-like masses.
\item The effective quarks are close to the mass shell such that the formation
  of additional quark-antiquark pairs is suppressed.
\end{itemize}

In such a scenario one can compute the projection of the density matrix $\rho$
of effective quarks onto hadron states
\begin{equation}
  N_h = \int \frac{d^3 P}{(2\pi)^3} \langle h; \mathbf{P} | \rho | h;
  \mathbf{P} \rangle
\end{equation}
where mesons and baryons are represented by their valence quarks.
This approach is called the instantaneous quark recombination formalism since
it happens suddenly, and it only conserves 3 out of 4
components of the energy-momentum vector. From the projection formula
one can derive a straight-forward overlap integral for the spectrum of mesons
(baryons are analogous) coalescing from partons $a$, $b$ 
\cite{Fries:2003kq,Fries:2008hs}
\begin{equation}
  \frac{dN_M}{d^3 P} = \sum_{a,b} \int \frac{d^3 R}{(2\pi)^3} \int \frac{d^3 q
    d^3 r}{(2\pi)^3} W_{ab} \left( \mathbf{R}-\frac{\mathbf{r}}{2},
  \frac{\mathbf{P}}{2}-q; \mathbf{R}+\frac{\mathbf{r}}{2},
  \frac{\mathbf{P}}{2}+q \right) \Phi_M (\mathbf{r},\mathbf{q})
\end{equation}
where $W_{ab}$ is the Wigner function of partons $a$ and $b$ in the fireball
and $\Phi_M$ is the Wigner function of meson $M$, and $\mathbf{r}$ and
$\mathbf{q}$ are the relative position and momentum of the two quarks.

From this equation different implementations have emerged, ranging from full
phase space overlap integrals (e.g.\ the model by Greco, Ko and Levai [GKL] 
\cite{Greco:2003xt,Greco:2003mm} ) to simplified schemes using 1-dimensional 
momentum integrals as in collinear factorization (e.g.\ the models by Fries,
M\"uller, Nonaka and Bass [FMNB] 
\cite{Fries:2003vb,Fries:2003kq,Nonaka:2003hx,Nonaka:2003ew}, and Hwa and 
Yang [HY] \cite{Hwa:2003bn,Hwa:2002tu}). The quark Wigner function is usually
approximated by the $n$-quark phase space distribution. The 
hadron Wigner functions are not known a priori
and are usually modeled according to simple guiding principles
(e.g.\ exclusive wave functions in the case of FMNB) with a few simple
parameters to fit. Indeed it turns out that applying this formalism to phase
space densities of thermalized quarks at large momenta, makes the results
very insensitive to details of the hadron Wigner function, since
\begin{equation}
  W_{ab} \sim e^{-\frac{P/2-q}{T}} e^{-\frac{P/2+q}{T}} = e^{-\frac{P}{T}} .
\end{equation}
independent of the relative momentum $\mathbf{q}$ in the hadron Wigner function.

Despite their differences in detail all instantaneous recombination models share
common benefits and shortcomings \cite{Fries:2008hs}:
\begin{itemize}
\item They violate energy conservation on the level of $M/P_T$ and $k_T/P_T$
    where $M$ and $k_T$ are hadron masses and intrinsic transverse momenta
    of quarks. Hence we can only expect them to provide reasonable results for
    large enough $P_T$, at least 1-2 GeV/$c$.
\item None of these models enforce quark number conservation.
  Rather quarks at lower momentum are seen as a fixed background.
  Since the description should be limited to a small part of phase space there
  is no problem of entropy conservation in these instantaneous models. 
\item Recombination does not make any \emph{a priori} predictions about the
  quark or quark gluon plasma phase itself. However, if hadron spectra are 
  experimentally measured one can fit quark distributions \emph{before} 
  hadronization as input for recombination models.
\item This immediately leads to the most stringent test for recombination
  model: the slew of different hadrons measured should be fit by only
  \emph{one} quark distribution as input. All recombination models do this
  remarkably well, including describing hadrons that clearly break from hydrodynamic
  behavior like the $\phi$ meson.
\end{itemize}
  
Instantaneous recombination models give access to fundamental parameters of the quark
phase at intermediate $P_T$  which is modeled as being close to thermal 
equilibrium so that the concepts of temperature and collective flow apply. 
The temperature, flow profile and the volume of the fireball (or more
precisely the hadronization hypersurface) are fit parameters.
Typical away-from-equilibrium deviations needed to fit the data are
modifications to the flow that make the elliptic flow saturate at intermediate 
$P_T$ \cite{Fries:2003kq}. This saturation in the
data --- in Fig.\ \ref{fig:3} seen for 1 GeV/$c < P_T/n < 2$ GeV/$c$ is a
clear indication that thermalization is no longer perfect at intermediate $P_T$.
Using these assumptions models do well describing the differences in
suppression between baryons and mesons, and describing hadron ratios at intermediate 
$P_T$, including the $\phi$ meson.
The available data is not sufficient to completely constrain the space-time 
dependence of the flow profile. A simple factorized ansatz is usually used in
which an asymmetry in momentum space $v_2^q$ is imprinted on the quark
phase. Note that this is different from a blastwave where there is a strong 
correlation in direction and magnitude between the position vector $\mathbf r$
of a fluid cell in the transverse plane, and the local flow vector $\mathbf v$.

With the factorized ansatz it is easy to show that \cite{Fries:2003kq}
\begin{equation}
  v_2^h (P_T) = n_h v_2^q \left( \frac{P_T}{n_h} \right)
\end{equation}
where $n_h$ is the number of valence quarks in hadron $h$.
This leads naturally to the experimentally observed scaling law and is seen as
a direct observation of quark degrees of freedom with collective flow. However, one needs
to be cautious about the assumptions used for the factorization ansatz for the
quark flow field. Indeed, using blastwave-like flow profiles as they naturally
emerge from hydrodynamics lead to modifications of the scaling law which can 
not be reconciled with experimental data
\cite{Pratt:2004zq,Molnar:2004rr}. This remains an unsatisfactory situation
to this date. Recombination models lead to quark number scaling using flow
profiles which are not consistent with hydrodynamic concepts. On the other hand, quark
number scaling in data is extremely robust and holds to a surprisingly large
accuracy, and other attempts to explain the scaling have mostly
failed. E.g.\ hadronic transport models can get similar scaling but the
overall size of the elliptic flow is too small. Recent attempts in
viscous hydrodynamics are more promising but need fine-tuning of viscous 
freeze-out distributions which is not under control \cite{Dusling:2009df}.

Instantaneous recombination models are often supported by calculations of
jet energy loss and fragmentation at large $P_T$. This opens the possibility
to introduce a jet-like component in the quark phase (either leading partons
or full jet showers) and to allow for recombination of quarks coming partly
from the bulk and partly from jets. This is known as soft-hard or
thermal-shower recombination. Such cross terms can be seen as a step toward
describing the modified hadronization of jets piercing a fireball. They also
introduce jet-like correlations at intermediate $P_T$ although the absolute
yields are dominated by hadrons from coalescence of bulk fireball quarks.
The farthest reaching approach is the HY model which introduces a quark
distribution at hadronization \cite{Hwa:2003ic,Hwa:2004zd}
\begin{equation}
  f (P_T) = f_{\mathrm{soft}}(P_T) + f_{\mathrm{shower}}(P_T)
\end{equation}
and applies quark recombination to all quarks equally such that mesons 
formed from two jet-shower quarks reproduce jet fragmentation, mesons from two 
soft quarks represent the usual recombination from the fireball, and the
cross-terms describe the modifications to jet fragmentation in the medium. 

\begin{figure}[t]
\centerline{
\includegraphics[width=12.0cm]{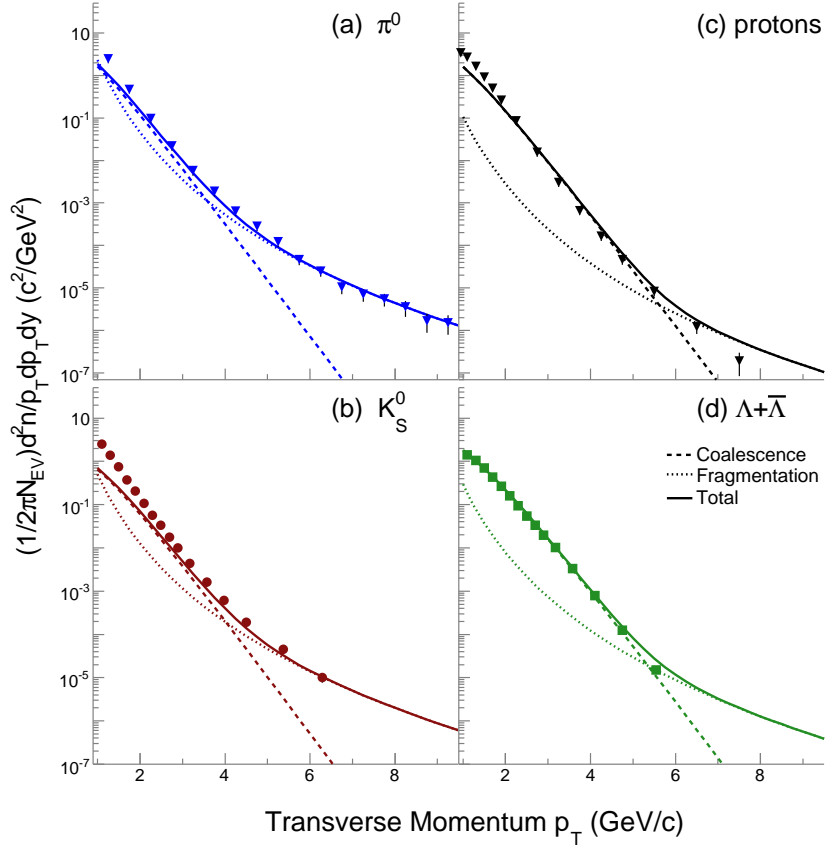}
%\hspace{-0.2cm}
%\includegraphics[width=8.0cm]{Rcp_summary.eps}
}
\caption{Spectra for neutral pions, kaons, protons and $\Lambda$ particles
calculated in the FMNB model including quark recombination and jet
fragmentation compared to data from RHIC \cite{Fries:2003kq}.} 
\label{fig:4}
\end{figure}

Fig.\ \ref{fig:4} shows spectra for four different hadrons calculated in the
FMNB model together with data from RHIC. The FMNB models uses quark
recombination at intermediate $P_T$ supplemented with a jet energy loss and
fragmentation calculation at high $P_T$. One can clearly see those two
domains with crossovers around 4 GeV/$c$ for mesons and around 6 GeV/$c$
for baryons. At small $P_T$ deviations from data start to occur due to missing
energy conservation and due to the mass mismatch for Goldstone bosons.

We want to conclude this section about recombination models by exploring ideas
to implement quark recombination for the bulk fireball at low transverse
momenta. Such a model clearly needs to enforce energy and momentum
conservation and should allow for the preservation of thermal and chemical
equilibrium. One approach is the resonance recombination model by Ravagli
and Rapp \cite{Ravagli:2007xx,Ravagli:2008rt}. It couples mesons as
resonances to a fixed background
of quarks which resembles the fireball just before hadronization.
Formation of mesons is governed by a Boltzmann equation with gain ($q+\bar q
\to M$) and loss ($M \to q+\bar q$) terms. To be precise
\begin{equation}\label{Boltzmann}
p^{\mu}\partial_{\mu}F_M(t,\mathbf{x},\mathbf{p})=-m\Gamma
F_M(t,\mathbf{x},\mathbf{p}) +p^0\beta(\mathbf{x}, \mathbf{p}),
\end{equation}
where $F_M(t,\mathbf{x}, \mathbf{p})$ denotes the phase space density of
the meson, and the gain term is given by
\begin{equation}\label{gainterm}
\beta(\mathbf{x},\mathbf{p})=\int\frac{d^3p_1d^3p_2}{(2\pi)^6}f_q(\mathbf{x},\mathbf{p})f_{\bar q}(\mathbf{x},\mathbf{p}) \sigma(s)v_{rel}(\mathbf{p}_1,\mathbf{p}_2)\delta^3(\mathbf{p} -\mathbf{p}_1
-\mathbf{p}_2),
\end{equation}
with the resonant cross section modeled by a relativistic
Breit-Wigner form:
\begin{equation}\label{Breit-Wigner}
\sigma(s)=g_{\sigma}\frac{4\pi}{k^2}\frac{(\Gamma
m)^2}{(s-m^2)+(\Gamma m)^2}.
\end{equation}

\begin{figure}[t]
\centerline{
\includegraphics[width=9.0cm]{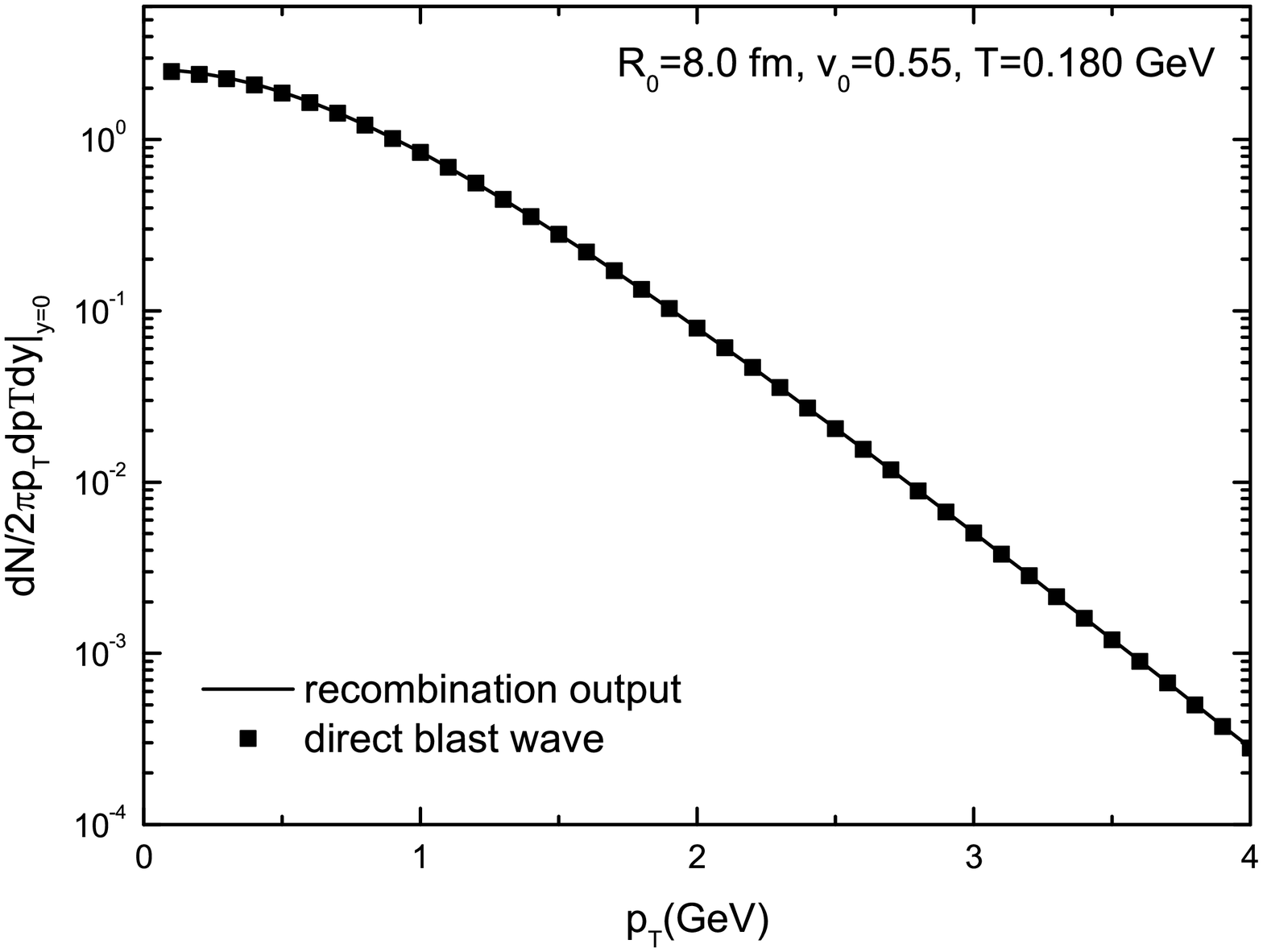}
\hspace{-0.2cm}
\includegraphics[width=9.0cm]{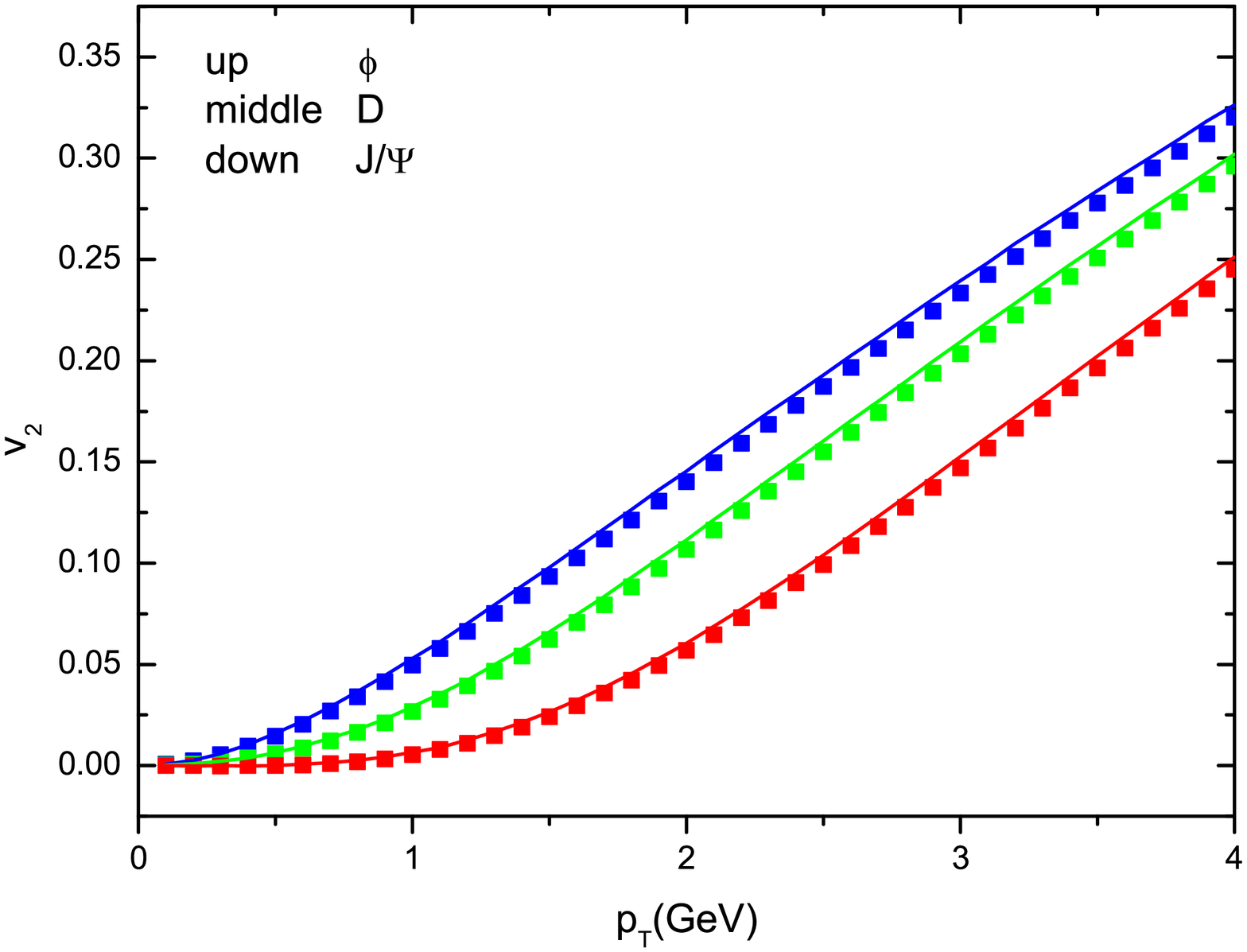}
}
\caption{Left panel: Spectrum of $\phi$ mesons calculated from a blastwave
  model, and from resonance recombination using quark phase space
  distributions from the
  same blastwave. Right panel: Elliptic flow $v_2$ for three different hadrons
  calculated from a blastwave and from resonance recombination using the same
  blastwave \cite{He:2010vw}.} 
\label{fig:5}
\end{figure}

The resonance recombination (RRM) approach has been criticized for treating
the quark phase as a static background, but note that the instantaneous
recombination formalism as well does not treat the quark phase explicitly. In
other words both formalisms do not address the question of confinement and 
could as well be applied to a theory with bound states but without confinement.
The big advantage of the RRM formalism is the conservation of momentum and 
energy, and the enforcement of detailed balance. Hence one can show that in 
the long-time limit a meson distribution recombining from a quark phase in 
local equilibrium is also in local equilibrium with the same temperature and 
collective flow velocity as the quark phase \cite{He:2010vw}.
This can be seen in Fig.\ \ref{fig:5} which shows spectra and elliptic flow of
hadrons obtained once directly from blastwave models with a temperature
equal to the critical temperature, and once from resonance recombination of
quarks where the quark phase space distributions have been determined from
the same blast wave at the critical temperature.

Because of its properties of energy conservation and detailed balance
resonance recombination is a first logical step to a comprehensive modeling of
quark recombination for the bulk fireball at low $P_T$. One of the more
impressive features is that it can produce negative values of $v_2$ (which in
blastwave and hydrodynamic models can occur at low $P_T$ for very heavy
particles) using quarks with strictly positive $v_2$. For instance this can
happen in $c\bar c \to J/\psi$ coalescence. This also indicates that there is
no simple scaling law for elliptic flow at low momenta, neither for quark
number nor for kinetic energy. 

Resonance recombination suggests that one can access the quark phase space
distributions at hadronization also as \emph{low} $P_T$. Note that the fitting
of quark phase space distributions to describe hadron data with instantaneous
coalescence at \emph{intermediate} $P_T$ works since hadronic rescattering for
hadrons with such large momentum is rather scarce and the measured
distribution resembles the spectrum at hadronization. This is not true for the
bulk fireball at low $P_T$ for which rescattering in the hadronic phase is
believed to be very important. However, one can analyze bulk hadrons which are
known to have very small cross sections and for which we have indication from data
that they freeze-out just below the critical temperature. Multi-strange
hadrons ($\phi$, $\Xi$, $\Omega$) are such hadrons. We can use them to fit 
quark distributions at low $P_T$ using resonance recombination, which has been
done in \cite{He:2010vw}.

\section{Open Questions and Outlook}

We have established quark recombination models to successfully explain key
features of hadron production in heavy ion collisions. We note that there is
no other natural mechanism to explain both the large yields of baryons, and
the quark number scaling law for elliptic flow at momenta of several GeV. 
However, recombination should not be seen as irreconcilable with hydrodynamics
and jet fragmentation. In fact we have proof of principle that quark
recombination can reproduce both fragmentation functions and local thermal 
equilibrium hadrons. If both regions joined smoothly at some intermediate
$P_T$ in the data recombination as a hadronization mechanism would not have 
come to our attention since fragmentation functions and a phase transition in the equation
of state in hydrodynamics would have taken care of hadronization.
However, at RHIC there seems to be a sufficiently large range of
momenta in which matter is not in local thermal equilibrium (note e.g.\ the
saturation of elliptic flow) but not at all in a ``dilute'' jet fragmentation
regime (note e.g.\ the large baryon/meson ratio). In this regime neither the
hydrodynamic nor the fragmentation concept are available and we have to resort
to some microscopic model which quark recombination supplies. Our discussion above
about successfully establishing either fragmentation or equilibrium distributions through
quark recombination shows that both at low and high momenta recombination is
compatible with the concepts available in those respective regions.

This gives recombination models a valuable place in heavy ion phenomenology.
The question whether quark recombination gives a direct glimpse of the parton 
phase, and, among other things, proves that collective flow is first carried
by partons needs further study. The fact that quark-number scaling laws have
only been proved
for unrealistic space-momentum correlations in the flow field is of concern.
On the other hand, kinetic energy scaling which has sometimes been advertised
as not compatible with quark recombination, is actually not at all related to it.
It is rather an effect of equilibrium and hydrodynamic flow. In Ref.\ \cite{He:2010vw}
it has been shown that it can be reproduced in the resonance recombination
model with some simple assumptions about freeze-out times. Kinetic energy
scaling seems to be somewhat accidental at RHIC and we should not be surprised
if it is not manifest at other collision energies. On the other hand,
quark-number scaling is a true test for quark recombination and we should see 
it hold at larger collision energies, as long as we can neglect rescattering 
in the hadronic phase. 

We also need to find a way to incorporate confinement and chiral symmetry 
breaking, i.e. hadron mass generation into recombination models. This would 
allow fully exclusive simulations that track the evolution of all quarks in a
sector of the fireball and hadronize them into hadrons obeying all
conservation laws and symmetries of QCD.

\section{Summary}

Quark recombination is an effective microscopic model for hadronization which works very
well to explain certain key features of the underlying dynamics. The basic 
concept is simple and has withstood dramatic
improvements. However, it has been shown that recombination can pass as
a microscopic model for both equilibrium hadronization and fragmentation.
Recombination models do not make predictions for the quark phase, but they can 
be used to extrapolate measured hadron data back to the time just before
hadronization.

I would like to thank Lanny Ray and Tom Trainor for suggesting this workshop
and for all their hard work organizing it. 
RJF is supported by CAREER Award PHY-0847538 from the U.S.\ National Science 
Foundation, by RIKEN/BNL and DOE grants DE-AC02-98CH10886, and
by the JET Collaboration under DOE grant DE-FG02-10ER41682.


\begin{thebibliography}{99}

\bibitem{whitepaper:05}
  J.~Adams {\it et al.}  [STAR Collaboration],
  %``Experimental and theoretical challenges in the search for the quark  gluon
  %plasma: The STAR collaboration's critical assessment of the  evidence from
  %RHIC collisions,''
  Nucl.\ Phys.\  A {\bf 757}, 102 (2005);
  %[arXiv:nucl-ex/0501009].
  %\bibitem{Adcox:2004mh}
  K.~Adcox {\it et al.}  [PHENIX Collaboration],
  %``Formation of dense partonic matter in relativistic nucleus nucleus
  %collisions at RHIC: Experimental evaluation by the PHENIX  collaboration,''
  Nucl.\ Phys.\  A {\bf 757}, 184 (2005).
  %[arXiv:nucl-ex/0410003].

\bibitem{Gyulassy:2004zy}
  M.~Gyulassy and L.~McLerran,
  %``New forms of QCD matter discovered at RHIC,''
  Nucl.\ Phys.\  A {\bf 750}, 30 (2005).


\bibitem{Becattini:2009fv}
  F.~Becattini and R.~Fries, in {\it Landolt-B\"ornstein: New Series},
  Vol.~I/23, ed.\ R.\ Stock,
  %``The QCD confinement transition: hadron formation,''
  [arXiv:0907.1031 [nucl-th]].
  %%CITATION = ARXIV:0907.1031;%%

%\cite{Fries:2008hs}
\bibitem{Fries:2008hs}
  R.~J.~Fries, V.~Greco and P.~Sorensen,
  %``Coalescence Models For Hadron Formation From Quark Gluon Plasma,''
  Ann.\ Rev.\ Nucl.\ Part.\ Sci.\  {\bf 58}, 177 (2008)
  [arXiv:0807.4939 [nucl-th]].
  %%CITATION = ARNUA,58,177;%%

%\cite{Fries:2006iv}
\bibitem{Fries:2006iv}
  R.~J.~Fries,
  %``Hadronization of dense partonic matter,''
  J.\ Phys.\ G {\bf 32}, S151 (2006)
  [arXiv:nucl-th/0609066].
  %%CITATION = JPHGB,G32,S151;%%

%\cite{Fries:2004ej}
\bibitem{Fries:2004ej}
  R.~J.~Fries,
  %``Recombination models,''
  J.\ Phys.\ G {\bf 30}, S853 (2004)
  [arXiv:nucl-th/0403036].
  %%CITATION = JPHGB,G30,S853;%%

\bibitem{Adler:2003kg} 
S.~S.~Adler {\it et al.}  [PHENIX Collaboration], 
%``Scaling properties of proton and anti-proton production in  s(NN)**(1/2) = 
%200-GeV Au + Au collisions,'' 
Phys.\ Rev.\ Lett.\  {\bf 91}, 172301 (2003).  
 
\bibitem{Adams:2003am} 
J.~Adams {\it et al.}  [STAR Collaboration], 
%``Particle type dependence of azimuthal anisotropy and nuclear modification of 
%particle production in Au + Au collisions at s(NN)**(1/2) = 200-GeV,'' 
Phys.\ Rev.\ Lett.\  {\bf 92}, 052302 (2004).  
 
\bibitem{Adams:2004bi} 
  J.~Adams {\it et al.}  [STAR Collaboration], 
  %``Azimuthal anisotropy in Au + Au collisions at s(NN)**(1/2) = 200-GeV,'' 
  Phys.\ Rev.\  C {\bf 72}, 014904 (2005).
  %[arXiv:nucl-ex/0409033]. 

\bibitem{Collins:1989gx}
  J.~C.~Collins, D.~E.~Soper and G.~Sterman,
  %``Factorization of Hard Processes in QCD,''
  Adv.\ Ser.\ Direct.\ High Energy Phys.\  {\bf 5}, 1 (1988),
  [arXiv:hep-ph/0409313].

%\cite{Albino:2008gy}
\bibitem{Albino:2008gy}
  S.~Albino,
  %``The hadronization of partons,''
  Rev.\ Mod.\ Phys.\  {\bf 82}, 2489 (2010)
  [arXiv:0810.4255 [hep-ph]].
  %%CITATION = RMPHA,82,2489;%%

\bibitem{Das:1977cp}
  K.~P.~Das and R.~C.~Hwa,
  %``Quark - Anti-Quark Recombination In The Fragmentation Region,''
  Phys.\ Lett.\  B {\bf 68}, 459 (1977); {\it Erratum-ibid.}  
  {\bf 73B}, 504 (1978).

\bibitem{Gupt:1983rq}
  C.~Gupt, R.~K.~Shivpuri, N.~S.~Verma and A.~P.~Sharma,
  %``Quark Anti-Quark Recombination In The Low Transverse Momentum Region,''
  Nuovo Cim.\  A {\bf 75}, 408 (1983)

\bibitem{Lopez:1984}
  J.~A.~Lopez, J.~C.~Parikh and P.~J.~Siemens,
  %``Testing QCD Plasma Formation By Pion Correlations In Relativistic Nuclear
  %Collisions,''
  Phys.\ Rev.\ Lett.\  {\bf 53}, 1216 (1984).

\bibitem{Biro:1994mp}
  T.~S.~Biro, P.~Levai and J.~Zimanyi,
  %``Alcor: A Dynamic Model For Hadronization,''
  Phys.\ Lett.\  B {\bf 347}, 6 (1995).

\bibitem{Aitala:1996hf}
  E.~M.~Aitala {\it et al.}  [E791 Collaboration],
  %``Asymmetries between the production of D+ and D- mesons from 500 GeV/c pi-
  %nucleon interactions as a function of xF and pt**2,''
  Phys.\ Lett.\  B {\bf 371}, 157 (1996).
  %[arXiv:hep-ex/9601001].

\bibitem{Braaten:2002yt}
  E.~Braaten, Y.~Jia and T.~Mehen,
  %``The leading particle effect from heavy-quark recombination,''
  Phys.\ Rev.\ Lett.\  {\bf 89}, 122002 (2002).
  %[arXiv:hep-ph/0205149].

\bibitem{Voloshin:2002wa}
  S.~A.~Voloshin,
  %``Anisotropic flow,''
  Nucl.\ Phys.\  A {\bf 715}, 379 (2003).
  %[arXiv:nucl-ex/0210014].

\bibitem{Hirano:2003pw}
  T.~Hirano and Y.~Nara,
  %``Interplay between soft and hard hadronic components for identified  hadrons
  %in relativistic heavy ion collisions at RHIC,''
  Phys.\ Rev.\  C {\bf 69}, 034908 (2004)
  [arXiv:nucl-th/0307015].
  %%CITATION = PHRVA,C69,034908;%%

\bibitem{Fries:2003kq} 
R.~J.~Fries, B.~Muller, C.~Nonaka and S.~A.~Bass, 
%``Hadron production in heavy ion collisions: Fragmentation and  recombination 
%from a dense parton phase,'' 
Phys.\ Rev.\ C {\bf 68}, 044902 (2003).  

\bibitem{Greco:2003xt} 
V.~Greco, C.~M.~Ko and P.~Levai, 
%``Parton coalescence and antiproton/pion anomaly at RHIC,'' 
Phys.\ Rev.\ Lett.\  {\bf 90}, 202302 (2003);  

\bibitem{Greco:2003mm} 
V.~Greco, C.~M.~Ko and P.~Levai, 
%``Parton coalescence at RHIC,'' 
Phys.\ Rev.\ C {\bf 68}, 034904 (2003).  

\bibitem{Fries:2003vb} 
R.~J.~Fries, B.~Muller, C.~Nonaka and S.~A.~Bass, 
%``Hadronization in heavy ion collisions: Recombination and fragmentation  of 
%partons,'' 
Phys.\ Rev.\ Lett.\  {\bf 90}, 202303 (2003).  

\bibitem{Nonaka:2003hx}
  C.~Nonaka, R.~J.~Fries and S.~A.~Bass,
  %``Elliptic flow of multi-strange particles: Fragmentation, recombination  and
  %hydrodynamics,''
  Phys.\ Lett.\  B {\bf 583}, 73 (2004)
  [arXiv:nucl-th/0308051].

\bibitem{Nonaka:2003ew}
  C.~Nonaka, B.~Muller, M.~Asakawa, S.~A.~Bass and R.~J.~Fries,
  %``Elliptic flow of resonances at RHIC: Probing final state interactions  and
  %the structure of resonances,''
  Phys.\ Rev.\  C {\bf 69}, 031902 (2004)
  [arXiv:nucl-th/0312081].

\bibitem{Hwa:2003bn} 
R.~C.~Hwa and C.~B.~Yang, 
%``Scaling distributions of quarks, mesons and proton for all p(T), energy  and %centrality,'' 
Phys.\ Rev.\ C {\bf 67}, 064902 (2003).

\bibitem{Hwa:2002tu}
  R.~C.~Hwa and C.~B.~Yang,
  %``Scaling behavior at high p(T) and the p/pi ratio,''
  Phys.\ Rev.\  C {\bf 67}, 034902 (2003).
  %[arXiv:nucl-th/0211010].

\bibitem{Pratt:2004zq} 
  S.~Pratt and S.~Pal, 
  %``Quark recombination and elliptic flow,'' 
  Nucl.\ Phys.\  A {\bf 749}, 268 (2005). 
  %[Phys.\ Rev.\  C {\bf 71}, 014905 (2005)] 
  %[arXiv:nucl-th/0409038]. 
 
\bibitem{Molnar:2004rr}
  D.~Molnar,
  %``Elliptic flow form quark coalescence: Mass ordering or quark number
  %scaling?,''
  {\it preprint} arXiv:nucl-th/0408044.

\bibitem{Dusling:2009df}
  K.~Dusling, G.~D.~Moore and D.~Teaney,
  %``Radiative energy loss and v2 spectra for viscous hydrodynamics,''
  Phys.\ Rev.\  C {\bf 81}, 034907 (2010)
  [arXiv:0909.0754 [nucl-th]].

\bibitem{Hwa:2003ic}
  R.~C.~Hwa and C.~B.~Yang,
  %``Recombination model for fragmentation: Parton shower distributions,''
  Phys.\ Rev.\  C {\bf 70}, 024904 (2004)
  [arXiv:hep-ph/0312271].

\bibitem{Hwa:2004zd}
  R.~C.~Hwa and C.~B.~Yang,
  %``Final-state interaction as the origin of the Cronin effect,''
  Phys.\ Rev.\ Lett.\  {\bf 93}, 082302 (2004);
  %[arXiv:nucl-th/0403001].
  %\bibitem{Hwa:2004in}
  R.~C.~Hwa, C.~B.~Yang and R.~J.~Fries,
  %``Forward production in d + Au collisions by parton recombination,''
  Phys.\ Rev.\  C {\bf 71}, 024902 (2005).
  %[arXiv:nucl-th/0410111].

\bibitem{Ravagli:2007xx}
  L.~Ravagli and R.~Rapp,
  %``Quark coalescence based on a transport equation,''
  Phys.\ Lett.\  B {\bf 655}, 126 (2007)
  [arXiv:0705.0021 [hep-ph]].

\bibitem{Ravagli:2008rt}
  L.~Ravagli, H.~van Hees and R.~Rapp,
  %``Resonance Recombination Model: A Dynamical Framework for Hadronization,''
  Phys.\ Rev.\  C {\bf 79}, 064902 (2009)
  [arXiv:0806.2055 [hep-ph]].
  %%CITATION = PHRVA,C79,064902;%%

\bibitem{He:2010vw}
  M.~He, R.~J.~Fries and R.~Rapp,
  %``Scaling of Elliptic Flow, Recombination and Sequential Freeze-Out of
  %Hadrons in Heavy-Ion Collisions,''
  Phys.\ Rev.\  C {\bf 82}, 034907 (2010)
  [arXiv:1006.1111 [nucl-th]].


\end{thebibliography}
\end{document}